\documentclass[amsmath,amssymb,aps,letterpaper,prd,twocolumn,longbibliography,10pt]{revtex4-1}
\pdfoutput=1
\usepackage{graphicx}
\usepackage{color}
\usepackage{comment}

\usepackage{subfigure}

\usepackage{color}

\makeatletter \let\@keywords\@empty \let\@subject\@empty
\providecommand{\keywords}[1]{\gdef\@keywords{#1}}
\providecommand{\subject}[1]{\gdef\@subject{#1}}
\def\thetitle{\@title}
\def\theauthor{\@author}
\def\thesubject{\@subject}
\def\thedate{\@date}
\def\thekeywords{\@keywords}
\makeatother
\AtBeginDocument{%
\hypersetup{pdftitle={\thetitle}}%
\hypersetup{pdfauthor={\theauthor}}%
\hypersetup{pdfsubject={\thesubject}}%
\hypersetup{pdfkeywords={\thekeywords}}%
}

\usepackage[bookmarks=true,hyperfigures=true]{hyperref}
\usepackage{fixmath}% makes latex math comply with international standards ISO 31-0:1992 to ISO 31-13:1992 (for capital greek letters)
\usepackage{mathtools}%for mathllap

\providecommand{\href}[2]{#2}

\let\oldbfseries=\bfseries
\let\oldmdseries=\mdseries
\let\oldnormalfont=\normalfont
\renewcommand{\bfseries}{\oldbfseries\boldmath}
\renewcommand{\mdseries}{\oldmdseries\unboldmath}
\renewcommand{\normalfont}{\oldnormalfont\unboldmath}

\makeatletter
\newlength{\apb@width}
\newcommand{\autoparbox}[2][c]{\settowidth{\apb@width}{#2}\parbox[#1]{\apb@width}{#2}}

\makeatother

\newcommand{\beq}{\begin{equation}}
\newcommand{\eeq}{\end{equation}}
\newcommand{\bea}{\begin{eqnarray}}
\newcommand{\eea}{\end{eqnarray}}

\newcommand{\nn}{\nonumber}

\newcommand{\CO}{\mathcal{O}}

\newcommand{\spa}{\ , \ \ }
\newcommand{\ds}{\displaystyle}

\begin{document}

\title{\Large Force-free electrodynamics near rotation axis of a Kerr black hole}

\author{Gianluca Grignani$^1$}%
 \email{grignani@pg.infn.it}
\author{Troels Harmark$^2$}%
 \email{harmark@nbi.ku.dk}
\author{Marta Orselli$^{1,2}$}%
  \email{marta.orselli@pg.infn.it}

\affiliation{%
$^1$Dipartimento di Fisica, Universit\`a di Perugia, I.N.F.N. Sezione di Perugia, \\ Via Pascoli, I-06123 Perugia, Italy \\
$^2$Niels Bohr Institute, Copenhagen University,  Blegdamsvej 17, DK-2100 Copenhagen \O{}, Denmark
}%

\begin{abstract}
Despite their potential importance for understanding astrophysical jets, physically realistic exact solutions for magnetospheres around Kerr black holes have not been found, even in the force-free approximation. Instead approximate analytical solutions such as the Blandford-Znajek  $\mbox{(split-)}$monopole, as well as numerical solutions, have been constructed. 
In this paper we consider a new approach to the analysis and construction of such magnetospheres. We consider force-free electrodynamics close to the rotation axis of a magnetosphere surrounding a Kerr black hole assuming axisymmetry. This is the region where the force-free approximation should work the best, and where the jets are located. We perform a systematic study of the asymptotic region with $\mbox{(split-)}$monopole, paraboloidal and vertical asymptotic behaviors. Imposing asymptotics similar to a $\mbox{(split-)}$monopole, 
we find under certain assumptions that demanding regularity at the rotation axis and the event horizon restricts solutions of the stream equation so much that it is not possible for a solution to be continuously connected to the static $\mbox{(split-)}$monopole around the Schwarzschild black hole in the limit where the rotation goes to zero. 
%One can see this result as a no-go theorem for this type of perturbative constructions. 
On the one hand, this result provides independent evidence to the issues discovered with the asymptotics of the Blandford-Znajek $\mbox{(split-)}$monopole in Ref.~\cite{Grignani:2018ntq}. On the other hand, we also point out possible caveats in our arguments that one could conceivably exploit to amend the perturbative construction of the Blandford-Znajek $\mbox{(split-)}$monopole.
\end{abstract}

\maketitle

%%%%%%%%%%%%%%%%%%%%%%%%%%%%%%%%%%%%%%%%%%%%%%%%%%%%%%%%%%%%%%%%%%%%%%%%%%%%%%%%
\section{Introduction}
%%%%%%%%%%%%%%%%%%%%%%%%%%%%%%%%%%%%%%%%%%%%%%%%%%%%%%%%%%%%%%%%%%%%%%%%%%%%%%%%

Astrophysical electromagnetic jets coming out of spinning compact objects, such as pulsars and active galactic nuclei, are among the most fascinating phenomena that we observe in our universe.
%
%One of the most fascinating phenomena that we observe in the sky is the astrophysical electromagnetic jets coming out of spinning objects, such as pulsars and active galactic nuclei ({\bf change sentence}). These highly collimated beams are among the most/highest energetic signals in our universe. 
Despite the many observations and the advanced numerical simulations, our current understanding of the physics of astrophysical 
%electromagnetic 
jets is incomplete.
%There are in fact several mysteries and unanswered questions about these very extravagant signals that need to be addressed. 
It is therefore important to construct an analytical model that, combined with detailed numerical simulations, could guide us in describing the physics of astrophysical jets and that could be used to improve our understanding of these phenomena.

In this paper we consider the magnetosphere of a spinning black hole. The magnetosphere, in concert with the plasma of an accretion disc surrounding the black hole, is responsible for the astrophysical jet. 
The mechanism that is thought to be responsible for the energy extraction and electromagnetic jet production from spinning black holes is the Blandford and Znajek (BZ) process \cite{Blandford:1977ds,Komissarov:2008yh,Lasota:2013kia,Beskin:2010iba,Gralla:2014yja}. It was proposed in the context of the so-called Force Free Electrodynamics (FFE) approximation. Within this approximation, the plasma is assumed to be in equilibrium with the electromagnetic field, making it magnetically dominated, and it is moreover assumed that one can neglect the energy density of the plasma.

%matter degrees of freedom do not play any dynamical role and can thus be ignored. The full stress energy tensor is well approximated by just the electromagnetic part which is therefore conserved. The FFE equations are
%
%\begin{equation}
%\label{FFE_eqs}
%D_{[\mu}F_{\nu \rho]}=0 \spa
%D_\mu  F^{\mu\nu} = - J^\nu 
%\end{equation}
%
%\begin{equation}
%\label{FFE2}
%F_{\mu\nu} J^\nu = 0 \spa J^\mu \neq 0 \,,
%\end{equation}
%
%with
%
%\begin{equation}
%F_{\mu \nu}=\partial_{\mu} A_{\nu}-\partial_{\nu}A_{\mu}
%\end{equation}
%
%where $A_\mu$ is the gauge potential, $F_{\mu\nu}$ the electromagnetic field strength and $J^\mu$ the current. These are basically the ordinary Maxwell's equations supplemented with the force free condition \eqref{FFE2}.

Despite the success in finding analytical solutions of the FFE Eqs.~\eqref{FFE_eqs} in flat space and for the magnetosphere surrounding Schwarzschild black holes (see for example \cite{Blandford:1977ds}), it has proven to be highly difficult to find solutions for the case of rotating black holes.
A proposal for constructing analytical solutions of FFE for Kerr magnetospheres was introduced in \cite{Blandford:1977ds} in the limit of slow rotation by means of a perturbative expansion in the angular momentum of the black hole, assuming the solutions are stationary and axisymmetric. This was considered for the so-called $\mbox{(split-)}$monopole and paraboloidal cases.
In \cite{Grignani:2018ntq} we examined the asymptotic behavior of the Blandford-Znajek $\mbox{(split-)}$monopole solution, finding that the first non-trivial perturbative correction does not behave as one would expect from a monopole.  

The asymptotic behavior of the electromagnetic field is crucial for our understanding of the magnetosphere. Otherwise, a solution that might superficially look like a monopole close to the black hole could diverge or change into a different type of magnetosphere, with different asymptotic behavior of the field lines. Astrophysical jets have a length scale much larger than the black hole itself. Hence, one needs to have control of the behavior of the electromagnetic field also far away from the black hole. However, it seems that the asymptotic behavior of the electromagnetic field has not been sufficiently addressed in the case of the BZ $\mbox{(split-)}$monopole solution and how this affects its perturbative construction in a small rotation limit. Our work  \cite{Grignani:2018ntq} was motivated by the works \cite{Tanabe:2008wm, Pan:2014bja} where the fourth order term in the perturbative expansion was considered and found in \cite{Tanabe:2008wm} to diverge. We found in \cite{Grignani:2018ntq}, using a matched asymptotic expansions approach,  that already at second order the perturbative solution of the FFE equations does not have the appropriate asymptotic behavior for a $\mbox{(split-)}$monopole. 

One of the goals of this paper is to understand better the perturbative construction of the BZ $\mbox{(split-)}$monopole solution. This we address using an entirely different approach to the problem. We find that, under the assumptions we consider in this paper, there are issues for the perturbative construction of the $\mbox{(split-)}$monopole solution which are special to that type of asymptotic behavior, and they do not persist for other types of asymptotic behaviors, such as magnetospheres with paraboloidal and vertical field lines \cite{Gralla:2015vta}.

The new approach that we introduce in this work is to consider a perturbative expansion of the analytical solutions of the FFE equations around the rotation axis of the Kerr black hole. We assume that the solution is stationary and axisymmetric, and thus with the same rotational symmetry as the Kerr black hole.
The FFE equations reduce to the so-called Stream equation for a flux function $\psi$ along with the current $I(\psi)$ and the angular velocity $\Omega(\psi)$ \cite{Gralla:2014yja}. In terms of the spherical coordinates $(r,\theta,\phi)$ of the Kerr-Schild coordinates  for the Kerr metric we expand around the rotation axis at $\theta=0$. The flux function is expanded as $\psi = \sum_{n=1}^\infty \theta^{2n} \psi_n(r)$ while $I(\psi)$ and $\Omega(\psi)$ are infinite series in powers of $\psi$ with $I(0)=0$. Given any three functions $\psi_1(r)$, $I(\psi)$ and $\Omega(\psi)$, one has a solution of the Stream equation that is regular at the rotation axis. We then impose regularity at the horizon and that the solution is well-behaved in the asymptotic region. We consider three possible asymptotic behaviors: monopole, paraboloidal and vertical.

The region around the rotation axis is particularly relevant for applications to astrophysical jets for two important reasons. Firstly, assuming stationarity and axisymmetry, the jets lie in this region very close to the rotation axis of the Kerr black hole. Secondly, this is the region in which the force-free approximation should work the best. One reason for this is that the velocities of the field lines are below the speed of light close to the rotation axis, hence it is possible for the plasma to follow the field lines as it should in the FFE approximation. Moreover, it is supported by GRMHD simulations where one finds that the energy density of the plasma is low in the region around the rotation axis \cite{Tchekhovskoy:2011zx}.

An advantage of our approach is that one can consider a finite rotation parameter $\alpha = J/(GM^2)$ of the Kerr black hole with angular momentum $J$ and mass $M$. 
For the monopole-type asymptotics, our approach with expanding the Stream-equation in powers of $\theta$ around the rotation axis yields the surprising result that $\psi_1(r)$ is highly constrained and in addition there are some constraints on $I(\psi)$ and $\Omega(\psi)$ as well. In particular, assuming that $\psi_1(r)$ does not go to zero in the asymptotic region, one gets that there are terms that go like powers of $(\log r)/r$ which one cannot set to zero for $\alpha > 0$. However, the perhaps most surprising result is that when performing an $\alpha\rightarrow 0$ limit there are terms which are divergent in $\alpha$. This is seen by performing a systematic expansion of $\psi_1(r)$ at large $r$ up to order $r^{-11}$. 

One of our results is that no matter the choice of $I(\psi)$ and $\Omega(\psi)$, there are always terms in the flux function that diverge for $\alpha\rightarrow 0$. It is important to note that this result relies on certain assumptions, such as that the $\theta\rightarrow 0$ and $r\rightarrow \infty$ limits of $\psi(r,\theta)$ commute and that the $\theta^2$ part of $\psi(r,\theta)$ in the $\theta\rightarrow 0$ limit is non-zero at the event horizon. 

If we instead consider paraboloidal and vertical type asymptotics, we find that, in the case of paraboloidal asymptotics one has to impose that $
\frac{\psi_n (r)}{r^n} $ is finite for $r\rightarrow \infty$
for all $n \geq 1$, while for the case of vertical asymptotics there are no constraints that need to be imposed, meaning that any choice of the function $\psi_1(r)$ solves the Stream equation for vertical type asymptotics.

The results of this paper strongly suggests that we need to revisit the BZ $\mbox{(split-)}$monopole solution, since its current perturbative construction seem to fail. We point out that a possible resolution would be to use a magnetic flux function $\psi(r,\theta)$ which is not smooth across the outer light surface.

The $\mbox{(split-)}$monopole solution has played a very important role as a simple model for astrophysical jets. Recently, it has been used to compute the expected power output relevant for the 
 radio loud/quiet dichotomy of active galactic nuclei \cite{Tchekhovskoy:2009ba}. Moreover, it has been used as a test case for many numerical simulations \cite{Komissarov:2001sjq,Komissarov:2004ms,McKinney:2004ka,McKinney:2006sc,Tchekhovskoy:2009ba,Palenzuela:2010xn,Contopoulos:2012py,Nathanail:2014aua,Mahlmann:2018ukr}. This poses the question of how to reconcile the numerical studies with the results of this paper. 
Finally, magnetospheres with split-monopole-like behavior of the field lines seem a good description of the Crab Pulsar, although this is a neutron star and hence our boundary conditions at the event horizon do not apply.

%%%%%%%%%%%%%%%%%%%%%%%%%%%%%%%%%%%%%%%%%%%%%%%%%%%%%%%%%%%%%%%%%%%%%%%%%%%%%%%%
\section{Force free electrodynamics around Kerr black hole}
%%%%%%%%%%%%%%%%%%%%%%%%%%%%%%%%%%%%%%%%%%%%%%%%%%%%%%%%%%%%%%%%%%%%%%%%%%%%%%%%

We begin by reviewing the equations for FFE in the background of a Kerr black hole following the expositions in \cite{McKinney:2004ka,Tanabe:2008wm,Pan:2015haa}.
The metric for the Kerr black hole in Kerr-Schild coordinates is
\begin{eqnarray}
&&ds^2=-\left(1-\frac{r_0 r}{\Sigma}\right)dt^2+\frac{2r_0 r}{\Sigma}dr dt+\left(1+\frac{r_0 r}{\Sigma}\right)dr^2 \cr
&&+\Sigma d\theta^2-\frac{2 a r_0 r \sin^2{\theta}}{\Sigma}d\phi dt 
-2a\left(1+\frac{r_0 r}{\Sigma}\right)\sin^2\theta d\phi dr \cr
&&+\left(\Delta+\frac{r_0 r(r^2+a^2)}{\Sigma}\right)\sin^2\theta d\phi^2 \,,
\label{metric}
\end{eqnarray}
with
\begin{equation}
\begin{array}{c} \ds
\Sigma = r^2+a^2 \cos^2 \theta \spa  \Delta=(r-r_+)(r-r_-) \,,  \\[2mm] \ds
  r_\pm = \frac{r_0}{2} (1 \pm \sqrt{1 - \alpha^2}) \spa a = \frac{r_0}{2} \alpha  \,.
\end{array}
\end{equation}
Here $\alpha$ is proportional to the angular momentum and is in the range $0\leq \alpha \leq 1$.
The Kerr black hole is stationary and axisymmetric.

The equations for force-free electrodynamics (FFE) are the Maxwell equations combined with conservation of the electromagnetic energy-momentum tensor,
\begin{equation}
\label{FFE_eqs}
\begin{array}{c}\ds
F_{\mu \nu}=\partial_{\mu} A_{\nu}-\partial_{\nu}A_{\mu} \spa
   F_{\mu\nu} D_\rho  F^{\nu\rho} = 0  \,,
\end{array}
\end{equation}
with the condition $D_\mu  F^{\mu\nu} \neq 0$ and
where $A_\mu$ is the gauge potential and $F_{\mu\nu}$ the electromagnetic field strength.
We assume the FFE configurations are stationary and axisymmetric around the same rotation axis as the Kerr black hole. Hence we can choose a gauge with $\partial_t A_\mu = \partial_\phi A_\mu = 0$. 
The magnetic flux $\psi(r,\theta)$ and the total electric current $I(r,\theta)$ are given by
\begin{equation}
\psi=A_\phi \spa I = \sqrt{-g} F^{\theta r}  \,.
\end{equation}
The angular velocity $\Omega$ of the magnetic field line satisfies $\partial_r A_t = - \Omega \, \partial_r \psi$ and $\partial_\theta A_t = - \Omega \, \partial_ \theta \psi$. It follows from this that
\begin{equation}
\label{int_omega}
\partial_r \Omega \partial_\theta \psi = \partial_\theta \Omega \partial_r \psi \,.
\end{equation}
Using the $t,\phi$ components of the second equation of \eqref{FFE_eqs} one gets furthermore 
\begin{equation}
\label{int_I}
\partial_r I \partial_\theta \psi = \partial_\theta I \partial_r \psi \,.
\end{equation}
The integrability conditions \eqref{int_omega}-\eqref{int_I} imply that $\Omega$ and $I$ are functions of $\psi$,
\begin{equation}
\label{intconds}
\Omega= \Omega(\psi) \spa I = I (\psi) \,.
\end{equation}
From \eqref{FFE_eqs} one finds furthermore the {\sl Stream equation}
\begin{equation}
\label{stream}
-\Omega\, \partial_\mu (\sqrt{-g}F^{t \mu})+\partial_\mu (\sqrt{-g}F^{\phi \mu})+F_{r \theta}\frac{dI}{d\psi} = 0 \,.
\end{equation}
Moreover, the toroidal magnetic field $B^{\phi}$ is 
\begin{equation}
\label{genBphi}
B^{\phi}= \frac{1}{\sqrt{-g}} F_{r\theta} =-\frac{I \Sigma+\big(\Omega r_0 r-a\big) \sin \theta \partial_{\theta}\psi}{\Delta \Sigma \sin^2\theta}.
\end{equation}
Finding a solution of the FFE equations corresponds to finding $\psi$, $\Omega$ and $I$ that solve the integrability conditions \eqref{intconds} and the Stream equation \eqref{stream}.

%%%%%%%%%%%%%%%%%%%%%%%%%%%%%%%%%%%%%%%%%%%%%%%%%%%%%%%%%%%%%%%%%%%%%%%%%%%%%%%%
\section{Expansion near rotation axis}
%%%%%%%%%%%%%%%%%%%%%%%%%%%%%%%%%%%%%%%%%%%%%%%%%%%%%%%%%%%%%%%%%%%%%%%%%

We now expand the fields $\psi$, $\Omega$ and $I$ around the rotation axis at $\theta=0$. 
We impose the physical requirement that there are no sources at the rotation axis. This means the one-form $A_\mu$ should be smooth at $\theta=0$. In particular, this requires $\psi=0$ at $\theta=0$. Moreover, $\psi$ can be expanded in even powers of $\theta$ for $\theta \rightarrow 0$,
\begin{equation}
\label{psi_expand}
\psi(r,\theta) = \sum_{n=1}^\infty \theta^{2n} \psi_n(r) \,.
\end{equation}
Indeed, a term with odd powers of $\theta$, or of the form $\theta^n (\log \theta)^m$, would mean that $A_\mu$ is not a smooth one-form at $\theta= 0$. Similarly, one finds that $\Omega(r,\theta)$ and $I(r,\theta)$ can be expanded in even powers of $\theta$. From regularity of $B^\phi$ in Eq.~\eqref{genBphi} at $\theta=0$ we find that $I=0$ at $\theta=0$. Using this with the integrability conditions \eqref{intconds} we find that $\Omega$ and $I$ have the following expansions around $\theta=0$
\begin{equation}
\label{omI_expand}
\Omega(\psi) = \frac{1}{r_0} \sum_{n=0}^\infty \omega_n \psi^n \spa I(\psi) = \frac{1}{r_0} \sum_{n=1}^\infty i_n \psi^n \,.
\end{equation}

The Stream equation \eqref{stream} can be written as
\begin{eqnarray}
\label{stream2}
\partial_\theta \left( \frac{\partial_\theta \psi}{\theta} \right) &=& (r^2+a^2) \left[ \partial_\theta ( a \sin^2 \theta B^\phi ) + \partial_r ( \sqrt{-g} F^{\phi r} ) \right] \nn \\ && - \partial_\theta \left[ \left( \frac{r^2+a^2}{\Sigma \sin \theta} - \frac{1}{\theta}\right)\partial_\theta \psi \right] \nn \\ && + (r^2+a^2) \left[ F_{r\theta} \frac{dI}{d\psi} - \Omega \partial_\mu ( \sqrt{-g} F^{t \mu} ) \right] .
\end{eqnarray}
Inserting the expansion \eqref{psi_expand}-\eqref{omI_expand} in this, one finds at order $\theta$ that the LHS is $8\psi_2(r)$ and the RHS is an expression involving $\psi_1(r)$ and its derivatives, and various constant. More generally, one can show from this at higher orders in $\theta$ that $\psi_k(r)$ for $k \geq 2$ is given by $\psi_1 (r)$ and its derivatives, along with the constants $\omega_n$, $i_n$, $r_0$ and $\alpha$. In other words, given any function $\psi_1(r)$, and any choice of constants $\omega_n$, $i_n$, $r_0$ and $\alpha$, one has a solution of the Stream equation \eqref{stream} that is regular at the rotation axis $\theta=0$. Below we impose regularity at the event horizon $r=r_+$ and in the asymptotic region $r\rightarrow \infty$, which give further conditions.

Notice that given a solution $\psi(r,\theta)$ one can generate a new solution by the rescaling
\begin{equation}
\label{rescaling}
\psi_n(r) \rightarrow \epsilon \, \psi_n (r) \spa \omega_n \rightarrow \epsilon^{-n} \omega_n \spa i_n \rightarrow \epsilon^{1-n} i_n \,,
\end{equation}
where $\epsilon$ is constant.

%%%%%%%%%%%%%%%%%%%%%%%%%%%%%%%%%%%%%%%%%%%%%%%%%%%%%%%%%%%%%%%%%%%%%%%%%%%%%%%%
\section{Horizon regularity}
%%%%%%%%%%%%%%%%%%%%%%%%%%%%%%%%%%%%%%%%%%%%%%%%%%%%%%%%%%%%%%%%%%%%%%%%%

At the event horizon   $r=r_+$ of the Kerr black hole we demand regularity of $\psi$ and $B^\phi$. Regularity of $\psi$ follows from regularity of $\psi_1(r)$ by Eq.~\eqref{stream2}. Regularity of $B^\phi$ then follows from
\begin{equation}
\label{horizon_bc}
\left( I \Sigma+\big(\Omega r_0 r_+-a\big) \sin \theta \partial_{\theta} \psi \right) \Big|_{r=r_+} = 0 \,,
\end{equation}
known as the Znajek condition \cite{1977MNRAS.179..457Z}.
At lowest order in $\theta$ this gives
\begin{equation}
\label{horizon_lowest}
\psi_1(r_+) = 0 \ \mbox{  or  }\  i_1 = \frac{2r_0 (a- r_+ \omega_0)}{a^2 +r_+^2}
\end{equation}

Suppose we have a solution with $\alpha \neq 0$ and $i_1 =\omega_0=0$. From \eqref{horizon_lowest} we see that $\psi_1(r_+)=0$. It is straightforward to prove recursively using \eqref{horizon_bc} that $\psi_n(r_+)=0$ for all $n \geq 1$. Using the Stream equation \eqref{stream2}, order by order in $\theta$, one can show this implies that the derivatives of $\psi_1(r)$ are zero at $r=r_+$
\footnote{This is true except for isolated values of $\alpha$ where some derivatives can be non-zero. However, since we are interested in solutions that can work for a continuous range of values of $\alpha$ we can ignore this subtlety.}.
Thus, we have shown that $\psi_1(r)=0$ and thereby $\psi(r,\theta)=0$ at least in a region that surrounds the horizon. But this means such a solution is not physically interesting as it does not interact with the black hole, and hence we discard it. 

Another case that will be relevant below is if $i_1 = \pm 2 \omega_0$ and $\psi_1(r_+) \neq 0$. Combining this with Eq.~\eqref{horizon_lowest} we find 
\begin{equation}
\label{omega0_fixed}
i_1 = 2\omega_0 = \frac{a}{r_+} \,.
\end{equation}
Note that $ \Omega_{\rm H}=a/(r_0r_+)$ is the angular velocity of the Kerr black hole so that for $\theta=0$ one has
\begin{equation}
\label{omega_0}
\Omega |_{\theta=0} = \frac{1}{2} \Omega_{\rm H} \,.
\end{equation}

%%%%%%%%%%%%%%%%%%%%%%%%%%%%%%%%%%%%%%%%%%%%%%%%%%%%%%%%%%%%%%%%%%%%%%%%%%%%%%%%
\section{Asymptotic region}
%%%%%%%%%%%%%%%%%%%%%%%%%%%%%%%%%%%%%%%%%%%%%%%%%%%%%%%%%%%%%%%%%%%%%%%%%

To understand a solution of the FFE equations in the background of the Kerr black hole, it is crucial that one knows the asymptotic behavior of the solution. For one thing, if one knows only the solution close to the event horizon the solution might diverge in the asymptotic region, rendering it meaningless. But also for its physical interpretation it is important. If the solution asymptotically behaves like a magnetic monopole we can interpret it as the black hole is interacting with a magnetic monopole. Moreover, knowing the asymptotic behavior of the magnetic field lines is crucial for application to the formation of astrophysical jets. 

Note that it is expected that far away from the black hole the FFE approximation should break down \cite{Tchekhovskoy:2009da}. Obviously, such a breakdown would affect the asymptotic region. We take here the attitude that we are solving a well-posed problem in the framework of FFE in the vicinity of a Kerr black hole, and that a careful analysis of this problem can be a good starting point for a more realistic model in which one can also include effects that go beyond FFE. This is further commented on in the conclusions.

We now define the types of asymptotic behaviors that we shall consider in this paper. 
We consider here for convenience the half of Kerr space-time defined by $0\leq \theta < \pi/2$. This is the part of space-time that is nearer to the rotation axis at $\theta=0$. 
We define the following three types of asymptotic behaviors of the flux function $\psi$
\begin{itemize}
\item Monopole-type asymptotics:
\begin{equation}
\label{mono_asympt}
\psi \  \mbox{finite for}\ \ 
r\rightarrow \infty \ \ \mbox{with} \ \ \theta = \mbox{fixed} \, .
\end{equation}
\item Paraboloidal-type asymptotics:
\begin{equation}
\label{para_asympt}
\psi \  \mbox{finite for}\ \ 
r\rightarrow \infty \ \ \mbox{with} \ \ r \theta^2 = \mbox{fixed} \, .
\end{equation}
\item Vertical-type asymptotics:
\begin{equation}
\label{vert_asympt}
\psi \  \mbox{finite for}\ \ 
r\rightarrow \infty \ \ \mbox{with} \ \ r \theta = \mbox{fixed} \, .
\end{equation}
\end{itemize}
We see that in all three cases we impose that the magnetic flux function $\psi$ should be finite when following the magnetic field lines towards the asymptotic region.
One can motivate these three types of asymptotics by considering the non-rotating limit of the Kerr black hole with $a=0$, {\sl i.e.} the Schwarzschild space-time. The monopole asymptotics generalize the asymptotic behavior of the non-rotating monopole given by
\begin{equation}
\label{static_monopole}
\psi = 1-\cos \theta \,,
\end{equation}
with $\Omega=I=0$. In this case the magnetic field lines are simply the radial curves. However, the monopole asymptotics apply as well to the Michel monopole \cite{1973ApJ...180L.133M} and the hyperbolic solution \cite{Gralla:2015vta}. 
The paraboloidal asymptotics generalize the asymptotic behavior of the paraboloidal solution 
\begin{equation}
\psi = r (1-\cos \theta) + r_0 (1+\cos \theta) (1-\log (1+\cos \theta) )  \,.
\end{equation}
This is clear from following the paraboloidal field lines for $r\rightarrow \infty$, since the leading term for large $r$ is $\psi \simeq \frac{1}{2} r \theta^2$. Finally, the vertical asymptotics generalize the asymptotic behavior of the non-rotating vertical solution
\begin{equation}
\psi = r^2 \sin^2 \theta \,,
\end{equation}
with $\Omega=I=0$ since for large $r$ the flux function is $\psi\simeq r^2 \theta^2$ when following the magnetic field lines. This also applies to the perturbative solution with vertical field lines around the Kerr black hole found in \cite{Pan:2014bja}.

Below in Section \ref{sec:LS} we comment on the relation to the outer light surface and whether the limits $\theta\rightarrow 0$ and $r\rightarrow \infty$ commute for the three different types of asymptotics.

%%%%%%%%%%%%%%%%%%%%%%%%%%%%%%%%%%%%%%%%%%%%%%%%%%%%%%%%%%%%%%%%%%%%%%%%%%%%%%%%
\section{Inner and outer light surfaces}
\label{sec:LS}
%%%%%%%%%%%%%%%%%%%%%%%%%%%%%%%%%%%%%%%%%%%%%%%%%%%%%%%%%%%%%%%%%%%%%%%%%

Consider the co-rotation vector field
\begin{equation}
\chi = \partial_t + \Omega (\psi) \partial_\phi
\end{equation}
This is co-rotating with the force-free magnetosphere. When $\chi$ is null, $\chi^2 = 0$, one has a {\sl light surface}. On this surface one would have to move with the speed of light to follow the magnetic field lines. For a configuration with $0 < \Omega(\psi) < \Omega_H$, there are two light surfaces, called the inner and outer light surfaces. We write the location of the inner light surface as $r=r_{\rm ILS}(\theta)$ and the outer light surface as $r=r_{\rm OLS}(\theta)$.
The outer light surface $r=r_{\rm OLS}(\theta)$ can be understood from considering a small but constant $\Omega(\psi)$. Then one has a cylindrical radius $1/\Omega(\psi)$ for which the magnetic lines move at the speed of light. Thus, the outer light surface moves in from infinity as one turns on $\Omega(\psi)$ from zero. Instead the existence of the inner light surface $r=r_{\rm ILS}(\theta)$ one can infer starting with  $\Omega(\psi)=0$ from the existence of the ergosphere in the Kerr black hole since that has $g_{tt}=0$. 
As shown in \cite{Gralla:2014yja}, the Stream equation \eqref{stream} is singular on the inner and outer light surfaces.

The inner light surface $r=r_{\rm ILS}(\theta)$ ends at the rotation axis, and can therefore be seen in a small $\theta$ expansion. Nevertheless, we can choose to ignore it in our analysis. The reason for this is that the only information that we carry from the event horizon to the asymptotic region consists of the two parameters $\omega_0$ and $i_1$. These two parameters are determined by the functions $\Omega(\psi)$ and $I(\psi)$ that are the same functions everywhere in the space-time, due to the integrability conditions \eqref{int_omega}-\eqref{int_I} that follows from the force-free equations \eqref{FFE_eqs}. In particular, $\Omega(\psi)$ and $I(\psi)$ are the same functions on both sides of the inner light surface. Thus, to summarize, the fact that we can relate the values of $\omega_0$ and $i_1$ between the event horizon, as studied above, and the asymptotic region, as studied below, is due to the integrability conditions, and this is therefore not affected by the singular behavior of the Stream equation \eqref{stream} at the inner light surface. Numerical evidence for this also follows from \cite{Komissarov:2001sjq,Komissarov:2004ms,McKinney:2004ka,McKinney:2006sc,Tchekhovskoy:2009ba,Palenzuela:2010xn,Contopoulos:2012py,Nathanail:2014aua,Mahlmann:2018ukr} since all the numerical simulations are seen to obey the conditions \eqref{omega0_fixed} and \eqref{omega_0}.

Regarding the outer light surface $r = r_{\rm OLS}(\theta)$, our small $\theta$ expansion means that we are staying within the {\sl inner region} defined by $r < r_{\rm OLS}(\theta)$ since we first take a small $\theta$ limit and only afterwards look at large $r$ when studying the asymptotics of the solution. 
However, for the monopole and paraboloidal-type asymptotics, Eqs.~\eqref{mono_asympt} and \eqref{para_asympt} we clearly go outside of this region, and enter the {\sl outer region} defined by $r > r_{\rm OLS}(\theta)$. Therefore, if one does not have a magnetic flux $\psi(r,\theta)$ which is smooth across the outer light surface, the asymptotics that we find by our method would not necessarily match the asymptotics outside the outer light surface $r = r_{\rm OLS}(\theta)$. Instead, one can think of the asymptotics that we are obtaining as that of the magnetic flux $\psi(r,\theta)$ in the inner region $r < r_{\rm OLS}(\theta)$ if one analytically extends $\psi(r,\theta)$ in this patch beyond the outer light surface. Thus, our conclusions concerning the asymptotics is relevant for this analytically extended version of the magnetic flux $\psi(r,\theta)$. Of course, in case one has smoothness of $\psi(r,\theta)$ across the outer light surface, then the analytical extension should be identical the magnetic flux $\psi(r,\theta)$ in the outer region as well.
We comment further on these points in the conclusions.

Another point regarding the asymptotic behaviors \eqref{mono_asympt}-\eqref{vert_asympt} of the flux function $\psi$ is the question of the commutativity of the two limits $\theta\rightarrow 0$ and $r \rightarrow \infty$. We are assuming these limits commute in the case of monopole-type asymptotics for $\psi(r,\theta)$ in the inner region $r < r_{\rm OLS}(\theta)$. This is indeed satisfied by all known solutions \cite{Blandford:1977ds,McKinney:2004ka,Tanabe:2008wm,Pan:2015haa,Gralla:2015vta}. If one imagines a magnetic flux $\psi(r,\theta)$ where these limits do not commute, this would mean that our conclusions on the monopole-type asymptotics do not apply.

%%%%%%%%%%%%%%%%%%%%%%%%%%%%%%%%%%%%%%%%%%%%%%%%%%%%%%%%%%%%%%%%%%%%%%%%%%%%%%%%
\section{Monopole-type asymptotics}
%%%%%%%%%%%%%%%%%%%%%%%%%%%%%%%%%%%%%%%%%%%%%%%%%%%%%%%%%%%%%%%%%%%%%%%%%

We consider here the case of monopole-type asymptotics \eqref{mono_asympt} of the flux function. 
If one computes the right-hand side of the Stream equation \eqref{stream2} for a generic function $\psi(r,\theta)$ with asymptotics \eqref{mono_asympt} then it goes like $r^2$ for large $r$ due to the terms
\begin{equation}
\label{div_terms}
(r^2+a^2) \left[ F_{r\theta} \frac{dI}{d\psi} - \Omega \partial_\mu ( \sqrt{-g} F^{t \mu} ) \right] \,,
\end{equation}
 in the last line of \eqref{stream2}. This is at odds with the left-hand side which is finite for $r\rightarrow \infty$. Thus, a generic solution of the Stream equation does not satisfy the correct asymptotics. To have the correct asymptotics, we should impose on a solution of the Stream equation that the terms \eqref{div_terms} are finite. Analyzing \eqref{div_terms} for monopole-type asymptotics \eqref{mono_asympt} we see that the asymptotically leading part of $\psi$, $I$ and $\Omega$, here denoted as $\psi_{\infty}$, $I_{\infty}$ and $\Omega_{\infty}$,  should obey
\begin{equation}
\label{psi_asympt}
\sin \theta \Omega_{\infty} \frac{d\psi_{\infty}}{d\theta} = \pm I_{\infty} \,.
\end{equation}
This is derived in \cite{Grignani:2018ntq}. However, while this is necessary, it is not a sufficient condition for the terms \eqref{div_terms} being finite for $r\rightarrow \infty$. Hence, one needs to impose further conditions on the functions $\psi$, $I$ and $\Omega$ that go beyond the leading asymptotic parts $\psi_{\infty}$, $I_{\infty}$ and $\Omega_{\infty}$.

%%%%%%%%%%%%%%%%%%%%%%%%%%%%%%%%%%%%%%%%%%%%%%%%%%%%%%%%%%%%%%%%%%%%%%%%%%%%%%%%
\subsection{Angular expansion}
%%%%%%%%%%%%%%%%%%%%%%%%%%%%%%%%%%%%%%%%%%%%%%%%%%%%%%%%%%%%%%%%%%%%%%%%%

We now analyze the consequences of the necessary conditions for having a monopole-type asymptotics \eqref{mono_asympt} for a solution of the Stream equation \eqref{stream2} that is expanded in $\theta$ around the rotation axis at $\theta=0$.

Assuming the angular expansion \eqref{psi_expand} with the monopole-type asymptotics \eqref{mono_asympt}, we need to impose
\begin{equation}
\label{asympt_cond_mono}
\psi_n (r) \  \mbox{finite} \ \ \mbox{for} \ \ r\rightarrow \infty \,,
\end{equation}
for all $n \geq 1$.
In particular for $\psi_1(r)$ we write  
\begin{equation}
\psi_1(r) =   f_0 + \sum_{n=1}^\infty \left( \frac{r_0}{r} \right)^n f_n(r)  \,,
\end{equation}
for large $r$ where we require $f_n(r)/r \rightarrow 0$ for $r\rightarrow \infty$.

For generic choices of $\psi_1(r)$, $\omega_n$, $i_n$, $r_0$ and $\alpha$ one sees from the last line of Eq.~\eqref{stream2} that $\psi_{n+1}(r)$ goes like $r^{2n}$ for $r\rightarrow\infty$. Thus, $\psi_{n+1}(r)$ is of the form
\begin{equation}
\psi_{n+1}(r) = \sum_{k=-2n}^\infty c_{n,k}(r) \left(\frac{r_0}{r} \right)^{k} \,,
\end{equation}
for large $r$ with $c_{n,k}(r)/r\rightarrow 0$ for $r\rightarrow \infty$ and $n\geq 1$. 
To obey \eqref{asympt_cond_mono} we should thus ensure that $c_{n,j}=0$ for $j=-2n,...,-1$ and that $c_{n,0}$ is finite for $r\rightarrow \infty$. The sufficient conditions for monopole-type asymptotics \eqref{mono_asympt} are therefore
\begin{equation}
\label{asympt_bc}
c_{n,-2}=0 \ \ , \  c_{n,-1} =0 \ \ , \ \frac{d}{dr} c_{n,0} =0  \ \ \mbox{for} \ \ n \geq 1 \,.
\end{equation}
These conditions are sufficient since it is clear that $c_{2,-4}=0$ if we have $c_{1,-2}=0$, and so forth. 

%%%%%%%%%%%%%%%%%%%%%%%%%%%%%%%%%%%%%%%%%%%%%%%%%%%%%%%%%%%%%%%%%%%%%%%%%%%%%%%%
\subsection{Starting point for analysis}
%%%%%%%%%%%%%%%%%%%%%%%%%%%%%%%%%%%%%%%%%%%%%%%%%%%%%%%%%%%%%%%%%%%%%%%%%

In the following we assume $\alpha \neq 0$ corresponding to the Kerr black hole. 

If one starts with $f_0=0$ it is straightforward to see that the Stream equation \eqref{stream2} combined with the conditions \eqref{asympt_bc} on the asymptotic behavior lead to $f_n(r)=0$ for $n \geq 1$. Hence this corresponds to a trivial solution.
Considering instead $f_0\neq 0$ and $\omega_0=0$ one gets from $c_{1,-2}=0$ that $i_1 =0$. Hence, as shown above, horizon regularity implies that this is a trivial solution as well.

We are thus led to consider solutions with 
\begin{equation}
f_0 \neq 0 \ \ \mbox{and}\ \ \omega_0 \neq 0 \,.
\end{equation}
Below we set $f_0=1$ since we can use the rescaling symmetry \eqref{rescaling} to set $f_0=1$ without loss of generality.

%%%%%%%%%%%%%%%%%%%%%%%%%%%%%%%%%%%%%%%%%%%%%%%%%%%%%%%%%%%%%%%%%%%%%%%%%%%%%%%%
\subsection{Solving for monopole-type asymptotics}
%%%%%%%%%%%%%%%%%%%%%%%%%%%%%%%%%%%%%%%%%%%%%%%%%%%%%%%%%%%%%%%%%%%%%%%%%

We consider here in detail the first few orders in the expansion in $\theta$.

Using the Stream equation \eqref{stream2} we compute for $\psi_2(r)$ 
\begin{equation}
\begin{array}{c} \ds
c_{1,-2} = - \frac{1}{8} ( i_1^2 - 4\omega_0^2 ) \,, \\[2mm] \ds c_{1,-1} = - \frac{1}{8} ( i_1^2 - 4\omega_0^2 ) ( 1 + f_1(r)) \,.
\end{array}
\end{equation}
Imposing \eqref{asympt_bc} for $n=1$ thus gives $i_1 =\pm 2 \omega_0$. This corresponds to the $\theta^2$ contribution in Eq.~\eqref{psi_asympt} when one expands that equation for $\theta\rightarrow 0$.
We choose here the solution
\begin{equation}
\label{thei1}
i_1 = 2\omega_0 \,,
\end{equation} 
since this is compatible with \eqref{omega0_fixed} (one can get the solution with the other choice of sign by making the replacement $i_k \rightarrow - i_k$ in all expressions). Thus, we have chosen the plus sign in Eq.~\eqref{psi_asympt}. With this, we find 
\begin{equation}
\label{c10}
c_{1,0}=-\frac{1}{12} \,.
\end{equation}

Using the Stream equation \eqref{stream2} we compute for $\psi_3(r)$ 
\begin{equation}
c_{2,-2} = - \frac{1}{4} \omega_0 \left( i_2 - 2\omega_1 + \frac{1}{3} \omega_0 - 2 c_{1,0} \omega_0 \right) \,.
\end{equation}
This can alternatively be obtained from \eqref{psi_asympt} at order $\theta^4$.
Using Eq.~\eqref{c10} and solving $c_{2,-2}=0$ for $i_2$ we get
\begin{equation}
\label{thei2}
i_2 = 2 \omega_1 - \frac{1}{2} \omega_0 \,.
\end{equation}
We compute now
\begin{equation}
\label{c2m1}
c_{2,-1}= - \frac{\omega_0^2}{48} \left( 12 \alpha \omega_0 - 6 r f'_1 + r^2 f''_1 \right) \,,
\end{equation}
where we used Eq.~\eqref{c10} and 
\begin{equation}
c_{1,1} = - \frac{1}{2} \omega_0  -\frac{1}{3} f_1 + \frac{1}{4} r f'_1 - \frac{1}{8} r^2 f''_1 \,.
\end{equation}
Note now that $c_{2,-1}=0$ is an ordinary differential equation (ODE) for $f_1(r)$. The general solution of this is
\begin{equation}
f_1(r) = b_1 + b_2 r^7  + \frac{12}{7} \alpha \omega_0 \log \frac{r}{r_0} \,,
\end{equation}
with integration constants $b_1$ and $b_2$. Since $f_1/r$ should go to zero for $r\rightarrow \infty$ we find $b_2=0$. Thus,
\begin{equation}
\label{thef1}
f_1(r) = \Upsilon  + \frac{12}{7} \alpha \omega_0 \log \frac{r}{r_0} \,,
\end{equation}
where $\Upsilon$ is an undetermined constant. At this point we have determined $c_{1,0}$, $c_{1,1}$ and $f_1$ in terms of $\alpha$, $\omega_0$ and $\Upsilon$.

We now turn to \eqref{asympt_bc} for $n=3$. Using the Stream equation \eqref{stream2} we compute
\begin{equation}
\label{thec3m2a}
c_{3,-2} =  -\frac{\omega_0}{12} \left(  2i_3 + \omega_1 - 4 \omega_2 + \frac{\omega_0}{45}  \right)  + \frac{2}{3} \omega_0^2 c_{2,0} \,,
\end{equation}
where we used \eqref{thei1}, \eqref{c10} and \eqref{thei2}. This can alternatively be obtained from \eqref{psi_asympt}  at order $\theta^6$. Notice that Eq.~\eqref{thec3m2a} means that requiring $c_{3,-2}=0$ implies $\frac{d}{dr}c_{3,-2}=0$ and hence $\frac{d}{dr} c_{2,0}=0$ as required by \eqref{asympt_bc} for $n=2$. 

We use now the Stream equation \eqref{stream2} to compute $c_{2,0}$. To do this, one needs in addition to compute $c_{1,1}$ and $c_{1,2}$ in terms of $f_1$ and $f_2$. Inserting \eqref{thef1} one finds an expression for $c_{2,0}$ in terms of $f_2$. Using this in \eqref{thec3m2a} we find
\begin{equation}
\label{thec3m2b}
\begin{array}{l}
 c_{3,-2} = -\frac{\omega_0}{12} \left(  2i_3 + \omega_1 - 4 \omega_2 \right)  
 \\[2mm]  
+ \frac{\omega_0^4}{72} \left(  2 \alpha^2 + 9 \Upsilon + 6 \Upsilon^2 - 8 f_2 + 8 r f_2' - r^2 f_2'' \right)
 \\[2mm] 
 + \frac{ \alpha  \omega_0^5}{42} \left( - 17- 7 \Upsilon  + (9+12 \Upsilon )\log \frac{r}{r_0} \right)
 \\[2mm] 
+ \alpha^2 \omega_0^6 \left( - \frac{2}{7} \log \frac{r}{r_0} + \frac{12}{49} \left( \log \frac{r}{r_0} \right)^2   \right) \,.
 \end{array}
\end{equation}
The equation $c_{3,-2}=0$ is an ODE for $f_2(r)$. The solution is
\begin{equation}
\label{thef2}
\begin{array}{l}\ds
f_2(r) = - \frac{3}{4\omega_0^3} ( 2i_3 + \omega_1 - 4 \omega_2)
+ \frac{\alpha^2}{4} + \frac{9}{8} \Upsilon  \\[3mm]\ds + \frac{3}{4} \Upsilon^2
+ \omega_0 \left( - \frac{165}{112} \alpha + \frac{39}{28} \alpha \Upsilon \right)
+ \frac{837}{392} \alpha^2 \omega_0^2 
\\[3mm] \ds
+ \alpha \omega_0 \left( \frac{27}{14} + \frac{18}{7} \Upsilon + \frac{117}{49} \alpha \omega_0  \right) \log \frac{r}{r_0}
\\[3mm] \ds
+ \frac{108}{49} \alpha^2 \omega_0^2 \left( \log \frac{r}{r_0} \right)^2 \,.
\end{array}
\end{equation}
In addition there is a homogenous piece of the form $b_1 r + b_2 r^8$ which is required to be zero in order to have $f_2/r\rightarrow 0$ for $r\rightarrow \infty$. 

We have now determined $f_1$, $f_2$, $c_{1,0}$, $c_{1,1}$, $c_{1,2}$ and $c_{2,0}$. Inserting these known functions into $c_{3,-1}$ computed from the Stream equation \eqref{stream2} we get
\begin{equation}
\begin{array}{l}\ds
c_{3,-1} = - \frac{\omega_0 \Upsilon}{4} ( 2 i_3 + \omega_1 - 4 \omega_2 ) + \frac{11}{504} \alpha \omega_0^3
\\[3mm]\ds
+ \left( \frac{2}{3} c_{2,1} - \frac{4}{135} \Upsilon + \frac{1}{4} \alpha \omega_1 \right) \omega_0^2 
\\[3mm] \ds
+ \alpha \omega_0^2 \left( \frac{12}{7} \omega_2 - \frac{3}{7} \omega_1 - \frac{16}{315} \omega_0 - \frac{6}{7} i_3 \right) \log \frac{r}{r_0} \,.
\end{array}
\end{equation} 
We use now the Stream equation \eqref{stream2} to get $c_{2,1}$ in terms of $f_3$, and in the process we use that one can determine $c_{1,3}$ in terms of $f_3$. This gives an expression for $c_{3,-1}$ in terms of $f_3$ as the only unknown function. The equation $c_{3,-1}=0$ then corresponds to an ODE for $f_3$, with solution
\begin{equation}
\label{thef3}
\begin{array}{l}\ds
f_3(r) =- \frac{1}{2\omega_0^3}(1+3\Upsilon) ( 2 i_3 + \omega_1 - 4\omega_2 )
\\[3mm] \ds
+ \frac{\alpha}{14\omega_0^2} ( 60 \omega_2 - 23 \omega_1 - 30 i_3 ) + \frac{3\alpha}{14\omega_0}  
\\[3mm]\ds
+ \frac{2\alpha^2}{3} + \frac{5}{4}\Upsilon + \frac{\alpha^2 \Upsilon}{4} + \frac{19 \Upsilon^2}{12} + \frac{\Upsilon^3}{2}
\\[3mm]\ds
+ \left( 2 \Upsilon^2 + \frac{379}{504} \Upsilon + \frac{\alpha^2}{3} - \frac{383}{168} \right) \alpha \omega_0
\\[3mm]\ds
+ \left( \frac{3445}{1512} + \frac{2467}{588} \Upsilon \right) \alpha^2 \omega_0^2 + \frac{28633}{12348} \alpha^3 \omega_0^3
\\[3mm]\ds
+ \left\{ - \frac{18\alpha}{7\omega_0^2} ( 2  i_3 + \omega_1 - 4\omega_2) 
 \right. \\[3mm]\ds 
+ \frac{\alpha\omega_0}{7} ( 15 + 3 \alpha^2 + 38 \Upsilon + 18 \Upsilon^2 )
\\[3mm]\ds \left.
+ \alpha^2 \omega_0^2  \left( \frac{379}{294} + \frac{48}{7} \Upsilon \right) + \frac{2467}{343} \alpha^3 \omega_0^3
\right\} \log \frac{r}{r_0}
\\[3mm] \ds
+ \frac{12}{49} \alpha^2 \omega_0^2 \left( 19 + 18 \Upsilon + 24 \alpha \omega_0   \right) \left( \log\frac{r}{r_0} \right)^2
\\[3mm] \ds
+ \frac{864}{343} \alpha^3 \omega_0^3 \left( \log\frac{r}{r_0} \right)^3 \,.
\end{array}
\end{equation} 
In addition there is a homogenous piece $b_1 r^2 + b_2 r^9$ that one should set to zero to have $f_3/r\rightarrow 0$ for $r\rightarrow \infty$.
We have now determined $f_1$, $f_2$, $f_3$, $c_{1,0}$, $c_{1,1}$, $c_{1,2}$, $c_{1,3}$, $c_{2,0}$ and $c_{2,1}$.

One can continue in this way solving systematically for the functions $f_k$ and $c_{n,j}$. The next step is to solve for $c_{3,0}$, $c_{2,2}$, $c_{1,4}$ and $f_4$ and subsequently  $c_{3,1}$, $c_{2,3}$, $c_{1,5}$ and $f_5$. One can continue this iteratively, solving $c_{n,0}$, $c_{n-1,2}$, $c_{n-2,4}$,..., $c_{2,2n-2}$ and $f_{2n-2}$ and subsequently $c_{n,1}$, $c_{n-1,3}$, $c_{n-2,5}$,..., $c_{2,2n-1}$ and $f_{2n-1}$. In all cases one ends up with a ODE for $f_k$ that one can solve. The general form of the solution is
\begin{equation}
f_n(r) = f_{n,0} + \sum_{j=1}^n f_{n,j} \left( \log \frac{r}{r_0} \right)^j \,.
\end{equation}
Here the coefficients $f_{n,j}$ are specific functions of $\alpha$, $\Upsilon$, $\omega_k$ and $i_k$ for $k \geq 0$.
Note that the homogenous piece is of the form $b_1 r^{n-1} + b_2 r^{n+6}$ so for $n \geq 2$ we need this to be zero to have $f_n(r)/r\rightarrow 0$ for $r\rightarrow \infty$. Note also that
\begin{equation}
f_{1,0} = \Upsilon \,.
\end{equation}
This is the only coefficient in the homogenous part of the ODE's for the $f_k(r)$ functions that can be non-zero.
One can read off more $f_{n,j}$ coefficients from the explicit solutions \eqref{thef1}, \eqref{thef2} and \eqref{thef3}.

We have used the iterative procedure explained above to solve for $f_1(r)$, $f_2(r)$, ..., $f_{11}(r)$ in terms of $\alpha$, $\Upsilon$, $\omega_k$ and $i_k$ for $k \geq 0$. The expressions for these are recorded in Mathematica files that we can send upon request. 
 With this, we have imposed the correct asymptotics on $\psi_2(r),... ,\psi_8(r)$.

While we have not found a general solution for the functions $f_n(r)$ we noticed a simple expression for the  coefficients of the highest powers of the logarithms 
\begin{equation}
f_{n,n} = (n+1) \left(\frac{6}{7} \alpha \omega_0 \right)^n \,.
\end{equation}
It would be highly interesting if one could use this as an alternative starting point for a recursive procedure.

We noticed the following general feature for the $f_{n,j}$ coefficients.
For $\omega_0 \rightarrow 0$ the leading term of $f_{n,j}$ behaves as
\begin{equation}
\label{fnj_omzero}
f_{n,j} \sim \left\{ \begin{array}{l} 
\omega_0^{-3[\frac{n}{2}]  + 5 \frac{j}{2}} \ \  \mbox{for}\  j \  \mbox{even}\,, \\ \omega_0^{1-3[\frac{n-1}{2}]  + 5 \frac{j-1}{2}}  \ \  \mbox{for}\  j \  \mbox{odd} \,.
\end{array} \right.
\end{equation}
where $[\frac{n}{2}]$ is the integer part of $\frac{n}{2}$.
We note also that $f_{n,j}$ has an overall factor of $\alpha^j$. These observations we shall use below.

Imposing in addition regularity on the event horizon one gets \eqref{omega0_fixed} (assuming $\psi_1(r_+)\neq 0$) stating that $\omega_0$ is half the angular velocity of the Kerr black hole
\begin{equation}
\label{theomzero}
\omega_0 =  \frac{1-\sqrt{1-\alpha^2}}{2\alpha}
\end{equation}

Using the above results one has a large family of solutions to the Stream equation \eqref{stream} parametrized by $\alpha$, $\Upsilon$, $\omega_k$ and $i_{k+2}$ for $k \geq 1$. These solutions are regular at the rotating axis $\theta=0$ and obey the monopole-type asymptotics \eqref{mono_asympt}. However, for a given such solution, it is not clear how it will behave if one approaches the event horizon. While the necessary condition \eqref{theomzero} is obeyed (assuming $\psi_1(r_+)\neq 0$) for horizon regularity, this is not sufficient to ensure regularity at the horizon. Thus, what we have is a large family of solutions with regularity at $\theta \rightarrow 0$ and $r\rightarrow \infty$, but not necessarily at $r \rightarrow r_+$. 

Moreover, one should also impose regularity at the inner and outer light surfaces. This is not guaranteed for an arbitrary member of the large family of solutions that we are considering, just as regularity at the event horizon is not guaranteed. It would be potentially interesting to consider this since one can use this to further restrict the family of solutions.

%%%%%%%%%%%%%%%%%%%%%%%%%%%%%%%%%%%%%%%%%%%%%%%%%%%%%%%%%%%%%%%%%%%%%%%%%%%%%%%%
\subsection{Taking the limit $\alpha\rightarrow 0$}
%%%%%%%%%%%%%%%%%%%%%%%%%%%%%%%%%%%%%%%%%%%%%%%%%%%%%%%%%%%%%%%%%%%%%%%%%

We now consider what happens in the limit $\alpha \rightarrow 0$. The question here is whether one can connect to a solution of the Stream equation in the background of a Schwarzschild black hole in this limit. 

We see from \eqref{theomzero} that $\omega_0$ goes like $\alpha$ for $\alpha\rightarrow 0$. Using now \eqref{fnj_omzero} together with the fact that $f_{n,j}$ has an overall factor of $\alpha^j$ in its explicit expression, we see that without fixing any of the parameters $\Upsilon$, $\omega_k$ and $i_{k+2}$ for $k \geq 1$, we have that the leading term of $f_{n,j}$ in the $\alpha \rightarrow 0$ limit goes like
\begin{equation}
\label{fnj_alpha}
f_{n,j} \sim \left\{ \begin{array}{l} 
\alpha^{-3[\frac{n}{2}]  + 7 \frac{j}{2}} \ \  \mbox{for}\  j \  \mbox{even} \\ \alpha^{2-3[\frac{n-1}{2}]  + 7 \frac{j-1}{2}}  \ \  \mbox{for}\  j \  \mbox{odd}
\end{array} \right.
\end{equation}
We now expand the parameters $f_{1,0}$, $\omega_k$ and $i_{k+2}$ for $k \geq 1$ in powers of $\alpha$. We can parametrize this as
\begin{equation}
\begin{array}{c}\ds
\omega_k = \sum_{m=0}^\infty \omega_{k,m}\alpha^m \spa i_k = \sum_{m=0}^\infty i_{k,m}\alpha^m
\\[4mm] \ds
\Upsilon = \sum_{m=0}^\infty u_m \alpha^m
\end{array}
\end{equation}
 From \eqref{fnj_alpha} we see that $f_{2,0} \sim \alpha^{-3}$. This means that there are three equations for the $\alpha$-expanded parameters that one needs to satisfy to avoid that $f_{2,0}$ diverges for $\alpha\rightarrow 0$. One could also consider $f_{11,0} \sim \alpha^{-15}$ giving 15 equations to satisfy. Considering in this way all the potentially divergent terms in $f_{n,j}$ for $n=0,1,...,11$ we see from \eqref{fnj_alpha} that one has in total 191 equations that need to be satisfied. 
 
 One finds several equations linear in $\omega_{1,0}$ and $i_{3,0}$ that implies $\omega_{1,0}=0$ and $i_{3,0}=2\omega_{2,0}$. With this, 51 of the 191 equations are satisfied. Next step one finds similarly $\omega_{2,0}=0$ and $i_{4,0}=2\omega_{3,0}$ (implying $i_{3,0}=0$). With this 16 more equations are satisfied. The next steps are that one finds $\omega_{3,0}=0$ and $i_{5,0}=2\omega_{4,0}$ (implying $i_{4,0}=0$) and then $\omega_{4,0}=0$ and $i_{6,0}=2\omega_{5,0}$ (implying $i_{5,0}=0$). After this we have 109 equations left to satisfy. One finds now equations that demand
\begin{equation}
i_{n+2} = 2 \omega_{n+1} - \frac{1}{2} \omega_n + \CO(\alpha^3)
\end{equation} 
for $n=1,2,3,4,5$. Subsequently, one finds an equation demanding $\omega_{5,0}=0$ (implying $i_{6,0}=0$). After this, there are 72 equations left to satisfy. One finds then equations for $\omega_{n,1}$ and $\omega_{n,2}$ with $n=2,3,4$ that are linear in the variables one solves for, but with the solutions being polynomial expressions in terms of $\omega_{1,1}$ and $\omega_{1,2}$. After this, one has 66 equations left to satisfy. Two of these equations are
\begin{equation}
\label{twopoly}
\begin{array}{rcl}
0&=&\frac{7797119}{165888}+\frac{815609}{13824}\omega_{1,1}-\frac{204899}{2592} \omega_{1,1}^2 + \omega_{1,1}^3 
\\[4mm]
0&=&-\frac{51308967545}{41803776}  - \frac{30557468503}{27869184} \omega_{1,1} \\[4mm] &&+\frac{14080362967}{5225472}\omega_{1,1}^2 
- \frac{412819}{567} \omega_{1,1}^3 + \omega_{1,1}^4 
\end{array}
\end{equation}
These two polynomials do not have any common roots. Hence it is not possible to satisfy both these equations.

The conclusion we can draw from the above consideration is the following. Suppose we are given a non-trivial smooth solution $\psi(r,\theta)$ of the Stream equation \eqref{stream} obeying the monopole-type asymptotic \eqref{mono_asympt} and with $\psi_1(r_+)\neq 0$. Then, under our assumptions, this cannot be connected to any of the solutions of the Stream equation \eqref{stream} in the background of the Schwarzschild black hole ({\sl i.e.} the $\alpha=0$ case). Thus, to generalize the known $\alpha=0$ solutions of the Stream equation \eqref{stream} with monopole-type asymptotics one would have to break one of our assumptions. 

In particular, the generalization of the static monopole solution \eqref{static_monopole} in the background of the Schwarzschild black hole, to a rotating monopole solution in the background of the Kerr black hole needs further analysis. Note that this also applies to the perturbative construction of the hyperbolic solution of \cite{Gralla:2015vta}.

%%%%%%%%%%%%%%%%%%%%%%%%%%%%%%%%%%%%%%%%%%%%%%%%%%%%%%%%%%%%%%%%%%%%%%%%%%%%%%%%
\section{Paraboloidal-type asymptotics}
%%%%%%%%%%%%%%%%%%%%%%%%%%%%%%%%%%%%%%%%%%%%%%%%%%%%%%%%%%%%%%%%%%%%%%%%%

We switch now to the case of paraboloidal-type asymptotics \eqref{para_asympt} of the flux function.

Consider first the left-hand side  of the Stream equation \eqref{stream2}. It is not difficult to see that $\partial_\theta ( \theta^{-1} \partial_\theta \psi )$ goes like $r^{3/2}$ in the limit \eqref{para_asympt}. Turning to the right-hand side of Eq.~\eqref{stream2}, the most divergent terms are written in Eq.~\eqref{div_terms}. The first term in \eqref{div_terms} is for large $r$
\begin{equation}
(r^2+a^2)F_{r\theta} \frac{dI}{d\psi} \simeq - \frac{r^2}{\theta} I \frac{dI}{d\psi} = - r^{5/2} \frac{I \frac{dI}{d\psi}}{\eta}
\end{equation}
where we defined
\begin{equation}
\eta \equiv \sqrt{r} \theta
\end{equation}
which is finite in the limit \eqref{para_asympt}. Thus, this term is more divergent than the left-hand side of Eq.~\eqref{stream2}. Considering the second term in \eqref{div_terms} we see that $(r^2 +a^2) (-\Omega) \partial_r ( \sqrt{-g} F^{tr} )$ goes like $r^{3/2}$ while 
\begin{equation}
(r^2 +a^2) (-\Omega) \partial_\theta ( \sqrt{-g} F^{t\theta} ) \simeq r^{5/2} \Omega \frac{d}{d\eta} \left( \eta \Omega \frac{d\psi}{d\eta} \right)
\end{equation}
We need that all the terms of order $r^{5/2}$ on the right-hand side of the Stream equation \eqref{stream2} cancel out. Hence, a necessary condition for having paraboloidal-type asymptotics is that 
\begin{equation}
\label{par_eq}
I_{\infty} \frac{dI_{\infty} }{d\psi_{\infty} } = \eta \Omega_{\infty}  \frac{d}{d\eta} \left( \eta \Omega_{\infty}  \frac{d\psi_{\infty} }{d\eta} \right)
\end{equation}
where $\psi_{\infty} (\eta) $, $\Omega_{\infty} (\eta) $ and $I_{\infty} (\eta)$ correspond to the asymptotic limits of $\psi$, $\Omega$ and $I$ in the limit \eqref{para_asympt}. 
If we define a new variable $\gamma$ by 
\begin{equation}
d\gamma = \frac{d\eta}{\eta \Omega_\infty}
\end{equation}
we can rewrite \eqref{par_eq} as
\begin{equation}
\frac{1}{2} \frac{d}{d\psi_\infty} I_\infty^2 = \frac{d^2 \psi_\infty}{d\gamma^2}
\end{equation}
Compute now
\begin{equation}
\frac{d}{d\gamma} \left( \frac{d\psi_\infty}{d\gamma} \right)^2 = 2 \frac{d\psi_\infty}{d\gamma} \frac{d^2 \psi_\infty}{d\gamma^2} = \frac{d}{d\gamma} I^2
\end{equation}
Integrating this equation we get $( \frac{d\psi_\infty}{d\gamma} )^2 = I^2$ using the boundary condition that $I_\infty \rightarrow 0$ for $\eta \rightarrow 0$. Therefore, we get
\begin{equation}
\label{para_condition}
\Omega_\infty \eta \frac{d\psi_\infty}{d\eta} = \pm I_\infty
\end{equation}
This is a necessary condition for having paraboloidal-type asymptotics \eqref{para_asympt}.

Assuming the angular expansion \eqref{psi_expand} for flux functions $\psi$ with paraboloidal-type asymptotics \eqref{para_asympt} we see that we need to impose
\begin{equation}
\label{asympt_cond_para}
\frac{\psi_n (r)}{r^n} \  \mbox{finite} \ \ \mbox{for} \ \ r\rightarrow \infty
\end{equation}
for all $n \geq 1$. One can check that this is satisfied provided \eqref{para_condition}. Thus, for the paraboloidal case it is sufficient to impose the condition \eqref{para_condition} on the asymptotic values of $\psi$, $\Omega$ and $I$. Hence, one does not get conditions on the subleading terms like for the case of monopole asymptotics.

%%%%%%%%%%%%%%%%%%%%%%%%%%%%%%%%%%%%%%%%%%%%%%%%%%%%%%%%%%%%%%%%%%%%%%%%%%%%%%%%
\section{Vertical-type asymptotics}
%%%%%%%%%%%%%%%%%%%%%%%%%%%%%%%%%%%%%%%%%%%%%%%%%%%%%%%%%%%%%%%%%%%%%%%%%

We turn finally to the case of vertical-type asymptotics \eqref{vert_asympt} of the flux function.
Considering the left-hand side of the Stream equation \eqref{stream2} one can see that $\partial_\theta ( \theta^{-1} \partial_\theta \psi )$ goes like $r^{3}$ in the limit \eqref{vert_asympt}. Considering now all the terms on the right-hand side, including the ones of Eq.~\eqref{div_terms}, one can check that they all go like $r^3$. Thus, the Stream equation is consistent with the vertical-type asymptotics \eqref{vert_asympt} of the flux function without the need of imposing any further conditions in the asymptotic region. 
Considering the angular expansion \eqref{psi_expand} for flux functions $\psi$ with vertical-type asymptotics \eqref{vert_asympt} we see it is equivalent to imposing
\begin{equation}
\label{asympt_cond_vert}
\frac{\psi_n (r)}{r^{2n}} \  \mbox{finite} \ \ \mbox{for} \ \ r\rightarrow \infty
\end{equation}
for all $n \geq 1$. One can readily check that this is always true given any function $\psi_1(r)$.

%%%%%%%%%%%%%%%%%%%%%%%%%%%%%%%%%%%%%%%%%%%%%%%%%%%%%%%%%%%%%%%%%%%%%%%%%%%%%%%%
\section{Conclusion and Outlook}
%%%%%%%%%%%%%%%%%%%%%%%%%%%%%%%%%%%%%%%%%%%%%%%%%%%%%%%%%%%%%%%%%%%%%%%%%%%%%%%%

In this paper we have considered an expansion of solutions of the Stream equation \eqref{stream} around the rotation axis at $\theta=0$ in the background of the Kerr black hole. This has enabled us to consider general features of solutions of the Stream equation for finite rotation parameter $\alpha$ of the Kerr black hole. Most importantly, we have analyzed solutions with monopole-type asymptotics \eqref{mono_asympt} for finite $\alpha$, with the following main results: 
\begin{itemize}
\item
We find that one has terms of the type
\begin{equation}
\label{surpriseterms}
\theta^2 (n+1) \left( \frac{6}{7} \alpha \omega_0 \frac{r_0}{r} \log \frac{r}{r_0} \right)^n  
\end{equation}
in the flux function $\psi$ that cannot be set to zero when requiring monopole-type asymptotics. These terms will induce similar non-zero terms at higher orders in $\theta$. 
\item In the $\alpha\rightarrow 0$ limit we find that any choice of solution of the Stream equation will diverge in negative powers of $\alpha$. Thus, it is not possible for any solution of the Stream equation with $\alpha$ non-zero that is regular at the horizon, at the rotation axis, and that obeys monopole-type asymptotics \eqref{mono_asympt} to be connected to a solution of the Stream equation at $\alpha=0$. This means that the perturbative construction of the Blandford-Znajek $\mbox{(split-)}$monopole is not possible, as it cannot be well-behaved asymptotically. 
\end{itemize}
These results rely on the following assumptions
\begin{itemize}
\item  $\psi_1(r_+)$ is non-zero.
\item The limits $\theta\rightarrow 0$ and $r\rightarrow \infty$ limits commute for the analytically extended $\psi(r,\theta)$ in the inner region $r < r_{\rm OLS}(\theta)$. 
\item $\psi(r,\theta)$ in the inner region $r < r_{\rm OLS}(\theta)$ obeys the monopole type asymptotics \eqref{mono_asympt}.
\item We have regularity of $\psi(r,\theta)$ at the rotation axis and at the event horizon.
\end{itemize}
Alternatively, one can also make the following stronger assumptions
\begin{itemize}
\item  $\psi_1(r_+)$ is non-zero.
\item $\psi(r,\theta)$ is smooth across the outer light surface $r = r_{\rm OLS}(\theta)$. 
\item The limits $\theta\rightarrow 0$ and $r\rightarrow \infty$ limits commute for $\psi(r,\theta)$. 
\item $\psi(r,\theta)$ obeys the monopole type asymptotics \eqref{mono_asympt}.
\item We have regularity of $\psi(r,\theta)$ at the rotation axis and at the event horizon.
\end{itemize}

We have also analyzed the conditions on the flux function $\psi$ in the case of paraboloidal and vertical-type asymptotics. For paraboloidal-type asymptotics \eqref{para_asympt} one has to impose the condition \eqref{para_condition} on the asymptotic values of $\psi$, $\Omega$ and $I$. Instead for vertical-type asymptotics \eqref{vert_asympt} there are no requirements on the flux function.

Our results reveal some unresolved issues in the understanding of the Blandford-Znajek $\mbox{(split-)}$monopole solution, at least in the way it has been constructed in the literature up to this point. Blandford-Znajek $\mbox{(split-)}$monopole has been used as an important analytical guide to the study of jet-physics for black holes since it was proposed in \cite{Blandford:1977ds}. 
Therefore it is important to clear up it's analytical realization, as the Blandford-Znajek $\mbox{(split-)}$monopole is used as a primary example of how one can have a magnetosphere around the Kerr black hole, and to demonstrate the power output of the black holes in a simple model (see for instance \cite{Tchekhovskoy:2009ba}). 
%Moreover, it is important to understand how it is possible to find the Blandford-Znajek $\mbox{(split-)}$monopole solution numerically, {\bf when it does not make sense analytically. It would be highly interesting to understand how to reconcile this.} 
As explained in the introduction, computing the power output for the split-monopole case has been important for the understanding of the radio loud/quiet dichotomy of active galactic nuclei \cite{Tchekhovskoy:2009ba}. 

%A possible explanation of some of the discrepancy between numerical and analytical results is that the most advanced numerical simulations include magnetohydrodynamics and thus go beyond force-free electrodynamics. While it seems that force-free electrodynamics should be valid near the rotation axis to a good approximation, it is conceivable that small deviations from force-free electrodynamics could change the boundary conditions at the horizon and/or asymptotically sufficiently in order for a modified version of the  Blandford-Znajek $\mbox{(split-)}$monopole solution to be possible. This is something we would like to pursue in the future. 

As we have explained in our paper, the inner light surface does not affect our analysis, even if it is ending at the rotation axis. This is because we only connect the analysis at the event horizon and in the asymptotic region through the two parameters $\omega_0$ and $i_1$. These two parameters are the same everywhere in the space-time due to the integrability conditions \eqref{int_omega}-\eqref{int_I} that ensure that $\Omega$ and $I$ are functions of $\psi$, and this statement is not affected by the singular behavior of the Stream equation \eqref{stream} at the inner light surface. However, since it is possibly to study the inner light surface for small $\theta$, demanding regularity of the magnetosphere at the inner light surface should provide further conditions than the ones we found by considering the event horizon and the asymptotic region. This would be highly interesting to consider. 

For the monopole-type asymptotics, we have assumed that the limits $\theta\rightarrow 0$ and $r\rightarrow \infty$ commute. As noted above, this is true for all known solutions \cite{Blandford:1977ds,McKinney:2004ka,Tanabe:2008wm,Pan:2015haa,Gralla:2015vta}. Nevertheless, given our above results it would be interesting to consider if one can evade our conclusions by using a magnetic flux function $\psi(r,\theta)$ for which these two limits do not commute.

Another, perhaps more promising avenue to pursue concerns the outer light surface. Our analysis is strictly concerned with the asymptotics in the inner region $r<r_{\rm OLS}(\theta)$, as explained above. 
However, it is conceivable that one could consider a magnetic flux function that is not smooth at the outer light surface. Assuming analyticity of $\psi(r,\theta)$ in the inner region, our results would still apply to there. Therefore, a possible resolution could by that the $\psi(r,\theta)$ in the inner region, analytically extended beyond the outer light surface, could have deviations from monopole-type asymptotics, and that this could be allowed by allowing for discontinuity of $\psi(r,\theta)$ across the outer light surface. This would be interesting to consider further.

In this paper we considered what happens in the limit $\alpha=J/(GM^2) \to 0$. There is however another interesting limit to explore. This is the limit $\alpha \to 1$, where the Kerr black hole is spinning close to extremality~\cite{Bardeen:1999px}. This is 
important also in view of the fact that actual astrophysical black holes giving rise to  electromagnetic jets are supposed to be rotating very fast, close to extremality~\cite{Tchekhovskoy:2011zx}. 
Examining this limit using the approach of this paper could provide important feedback when comparing the analytical results with numerical simulations. The Stream equation for the near horizon geometry of extremal Kerr magnetosphere, the so-called NHEK geometry, has received considerable attention in the literature \cite{Lupsasca:2014pfa,Lupsasca:2014hua,Gralla:2016jfc,Compere:2015pja} and it would be interesting to study the solutions found in the NHEK geometry using our small $\theta$ expansion.  
In fact, FFE exact solutions in the NHEK limit can be very important as starting points for a perturbative expansion around extremality in the same spirit as FFE exact solutions in a Schwarzschild background, such as the split-monopole, have been used as starting point for an expansion in small  $\alpha$, $ i.e.$ small angular momentum \cite{Blandford:1977ds}. 
However, note that the highest possible spin of astrophysical black holes has been estimated to be around $\alpha=0.998$ using thin accretion disks models \cite{Thorne:1974ve}. Therefore, the exact $\alpha = 1$ NHEK magnetosphere solutions might not suffice even for an approximate description.
In \cite{Bredberg:2009pv} it was considered another limit to go near the horizon in a non-extremal case, to obtain the so called near-NHEK geometry, where the spinning parameter $\alpha$ is near 1, but not exactly 1. This geometry should describe black holes near extremality, like the spinning black holes that give rise to the most energetic astrophysical jets. This is another interesting avenue to pursue using our new approach.

%%%%%%%%%%%%%%%%%%%%%%%%%%%%%%%%%%%%%%%%%%%%%%%%%%%%%%%%%%%%%%%%%%%%%%%%%%%%%%%%
\begin{acknowledgments}
%\paragraph{Acknowledgements.}
%%%%%%%%%%%%%%%%%%%%%%%%%%%%%%%%%%%%%%%%%%%%%%%%%%%%%%%%%%%%%%%%%%%%%%%%%%%%%%%%
T.~H.~acknowledges support from the Independent Research Fund Denmark grant number DFF-6108-00340 ``Towards a deeper understanding of black holes with non-relativistic holography". G.~G. and M.~O.~acknowledge support from the project ``Black holes, neutron stars and gravitational waves" financed by Fondo Ricerca di Base 2018 of the University of Perugia. T.~H.~thanks Perugia University and G.~G. and M.~O.~thank Niels Bohr Institute for hospitality.
\end{acknowledgments}

%\bibliography{mybib}


\end{document}